\documentclass[preprint,prd,noshowpacs,nofootinbib]{revtex4}
\usepackage{amssymb}
\usepackage{amsmath}
\usepackage{graphicx}

\begin{document}

\title{The time-dependent Aharonov-Casher effect}

\author{Douglas Singleton}
\email{dougs@csufresno.edu}
\affiliation{Department of Physics, California State University Fresno, Fresno, CA 93740-8031, USA \\
and \\
ICTP South American Institute for Fundamental Research,
UNESP - Univ. Estadual Paulista
Rua Dr. Bento T. Ferraz 271, 01140-070, S{\~a}o Paulo, SP, Brasil}

\author{Jaryd Ulbricht}
\email{julbrich@ucsc.edu}
\affiliation{Physics Department, University of California Santa Cruz, Santa Cruz, CA 95064 \\
and \\
Department of Physics, California State University Fresno, Fresno, CA 93740-8031, USA}

\date{\today}

\begin{abstract}
In this paper we give a covariant expression for Aharonov-Casher phase. This expression is a combination of the
canonical electric field, Aharonov-Casher phase plus a magnetic field phase shift. We use this covariant expression 
for the Aharonov-Casher phase to investigate the case of a neutral particle with a non-zero magnetic moment moving in 
the {\it time dependent} electric and magnetic fields of a plane electromagnetic wave background. We focus on the case 
where the magnetic moment of the particle is oriented so that both the electric and magnetic field 
lead to non-zero phases, and we look at the interplay between these electric and magnetic phases.    
\end{abstract}

\maketitle
\section{Introduction}
The Aharonov-Bohm (AB) effect \cite{AB, ES} and its experimental confirmation \cite{chambers, tonomura} are an
important consequence of combining quantum mechanics with gauge theories. The basic AB set-up is the quantum mechanical 
two-slit experiment for charged particles but with a magnetic-flux-carrying infinite solenoid placed between the 
two slits. Even though there is no magnetic field outside the solenoid, there is a non-vanishing vector potential, 
which leads to a shift in the interference pattern formed by the charged particles going through the slits. 
However, this shift in the interference pattern depends only on the magnetic flux carried inside the solenoid. This
gives the AB effect a non-local character since the phase of the particle is influenced by the magnetic field, which is
zero at the location of the particle. 

The closely related Aharonov-Casher (AC) effect \cite{AC, anadan} is also a modified version of the quantum
mechanical two-slit experiment, but using neutral particles with a non-vanishing magnetic moment
traveling through the slits with a line of charge placed between the slits. One again finds a quantum phase 
shift due to the presence of the electric field coming from the line of charge. This phase shift leads to an
observable shift in the interference pattern \cite{cimmino} \footnote{However, in the experiment carried out in
\cite{cimmino} the observed phase shift was 50{\%} larger than the theoretical prediction}. 

The AC effect can be seen as the dual of the AB effect in the following sense: In the AB set-up one has a
magnetic flux carrying solenoid, which can be pictured as a line of magnetic dipoles laid end-to-end, with
electrically charged particles moving around the solenoid. A duality transformation exchanges the magnetic dipoles
with the electric charge so that for the AC set-up one has a line of charges with neutral, magnetic dipoles moving 
around this line of charge. In both cases the particles pick up an additional quantum phase which manifests itself as 
a shift in the usual two-slit interference pattern.     

Both the AB and AC effects are topological since they depend on the non-simply connected nature of each
set-up. The infinite solenoid or the infinite line of charge exclude some region of the space-time
so that the space is not simply connected (for more discussion on this point see \cite{ryder}). Further the 
AB effect has been connected to Berry Phases \cite{berry} which are general ``geometric" phases that 
arise in certain quantum systems.

While the time-independent AB and AC effects have been extensively studied both theoretically and experimentally,
the same does not hold for the time-dependent cases when the magnetic flux in the solenoid or the
electric charge on the wire are time-dependent. The theoretical predictions as to the outcome of a time-dependent AB
experiment have been studied in \cite{chiao} \cite{singleton} \cite{singleton2}. There is some disagreement,
theoretically, as to the outcome of the time-dependent AB effect. The authors of \cite{chiao} predict that 
the interference pattern should shift with time according to the time dependence of the magnetic flux, whereas
the authors of \cite{singleton} \cite{singleton2} find a cancellation between the usual magnetic AB phase shift
and the additional phase shift that arises due to the electric field that exists in the time-dependent case. 
The one experiment performed to date on the time dependent AB effect \cite{chentsov, ageev} did find no
shifting of the interference pattern, but the parameters used in the set-up did not allow one to definitely 
conclude which prediction -- a time dependent shifting of the interference pattern or little/no time dependent
shifting of the interference pattern -- is correct. Thus additional experiments are called for in the
case of the time-dependent AB effect.

In this paper we write down a covariant version of the AC phase shift, and then use this expression to study
the time-dependent AC effect. The covariant expression which we find for the Aharonov-Casher phase has some
relationship to the geometrical phases obtained in \cite{casana} from Lorentz-violating terms from fermion interaction from
certain Standard Model extensions. 
\section{Covariant expression for Aharonov-Casher Phase}
\label{covariant}
In this section we begin by reviewing the derivation of the time-independent Aharonov-Casher phase \cite{AC} and also the derivation
of the Bernstein phase \cite{bernstein} which is also known as the scalar Aharonov-Bohm phase. \footnote{The scalar AB effect 
can refer to two things: (i) A phase shift due to the electric scalar potential, $ e \phi$, which is also called the
electric AB effect, (ii) or a phase shift due to the scalar potential ${\vec \mu} \cdot {\vec B}$. In this paper the scalar
AB effect means version (ii).} We use these two time-independent phases to give a motivation for our proposed 
expression for the covariant Aharonov-Casher phase, and then we give a more rigorous
derivation of this covariant expression for the Aharonov-Casher phase.   

In the original AC proposal \cite{AC} (see also \cite{anadan} for an earlier, closely related study)
a neutral, spin-$\frac{1}{2}$ particle was taken to move through an electric field. In reference \cite{AC}
the neutral, spin-$\frac{1}{2}$ particle was taken to be a neutron and the electric field was taken to be 
that of an infinite line of charge. The non-relativistic Hamiltonian for this system is   
\begin{equation}
\label{Hnr}
H_{NR} = \frac{1}{2m} \left( {\vec p} - {\vec E} \times {\vec \mu}  \right) ^2 - \frac{\mu ^2 E^2}{m} ~,
\end{equation}
where ${\vec p} = -i \nabla$ is the momentum operator and we have set $\hbar =1$. Also the vector magnetic 
moment is given as ${\vec \mu} = \mu {\vec s}$ (where the spin of the particle is ${\vec s} = \frac{1}{2} {\vec \sigma}$ 
with ${\vec \sigma}$ the standard Pauli matrices) and  
$E = | {\vec E} |$. The last term, $\frac{\mu ^2 E^2}{m}$, is negligible if $\mu E \ll m v$. The main
point to note about \eqref{Hnr} is that one can define an effective momentum operator
${\vec p} \rightarrow {\vec p} - {\vec E} \times {\vec \mu}$ analogously to the minimal coupling
definition of the momentum operator for a particle with non-zero electric charge $e$ 
i.e. ${\vec p} \rightarrow {\vec p} - e {\vec A}$ with ${\vec A}$ being the 3-vector potential. 
This operator, ${\vec p} - e {\vec A}$, results in the additional AB phase for a charged
particle moving along a path $L$ of the form
\begin{equation}
\label{phase-AB}
\Delta \alpha _{AB} = - e \int _L {\vec A} \cdot d {\vec r} ~.
\end{equation}
In the same way the operator, ${\vec p} - {\vec E} \times {\vec \mu}$, from \eqref{Hnr} results in an
additional AC phase for neutral particle with a magnetic moment moving along a path $L$ 
\begin{equation}
\label{phase-AC}
\Delta \alpha _{AC} = - \int _L ( {\vec E} \times {\vec \mu} )  \cdot d {\vec r} =
- \mu \int _L ( {\vec E} \times {\vec s} )  \cdot d {\vec r}~,
\end{equation}

Next we move on to the Bernstein phase \cite{bernstein} or the scalar AB effect. When a neutral particle with a magnetic 
momentum passes through a magnetic field the non-relativistic Hamiltonian for such a system is  
\begin{equation}
\label{Hnr-2}
H_{NR} = \frac{1}{2m} {\vec p} ~  ^2  - {\vec \mu} \cdot {\vec B}~.
\end{equation}
It was shown by Bernstein \cite{bernstein}  that this Hamiltonian also results in a quantum phase shift if 
the magnetic moment, ${\vec \mu}$, moves through a magnetic field, ${\vec B}$. The additional phase picked up 
due to the ${\vec \mu} \cdot {\vec B}$ term in \eqref{Hnr-2} can be written as
\begin{equation}
\label{phase-B}
\Delta \alpha _{B} =   \int _L ( {\vec \mu} \cdot {\vec B}  ) dt = 
 \mu \int _L ( {\vec s} \cdot {\vec B}  ) \frac{dx}{v} ~,
\end{equation}
where in the second integral the time integration has been turned into a space integration via
$v= \frac{dx}{dt}$ where $v$ is the velocity of the particle. This is done to make the connection
with the form of the Bernstein phase as given in the original paper \cite{bernstein}. However,
the main thing to note is that $\Delta \alpha _{AC}$ involves a spatial integration while $\Delta \alpha _B$  
involves a time integration. This previews the eventual combination of these two phase shifts into a covariant 
space-time integral. The magnetic moment in \eqref{phase-B} has again been split as ${\vec \mu} = \mu {\vec s}$. 
This phase shift given in \eqref{phase-B} was experimentally observed by Werner {\it et al.} \cite{werner}.

The phase shifts due to the electric field (the Aharonov-Casher effect) and magnetic field 
(the Bernstein effect) as written in equations \eqref{phase-AC} and \eqref{phase-B} are in 
non-covariant, 3-vector notation. Here we suggest a 4-vector, covariant  expression for 
the AC effect (which also therefore includes the Bernstein effect). The phase shifts in 
\eqref{phase-AC} and \eqref{phase-B} can be combined and written covariantly via the expression
\begin{equation}
\label{phase-ACB}
\Delta \alpha _{AC-B} =  \mu \int {\cal F} _{\mu \nu} S^\nu dx ^\mu ~,
\end{equation}
where $\mu$ is the magnitude of the magnetic moment defined previously. The term ${\cal F} _{\mu \nu}$ is the
dual Faraday field strength tensor ${\cal F} _{\mu \nu} = \frac{1}{2} \epsilon _{\mu \nu \alpha \beta} F^{\alpha \beta}$
with $\epsilon _{\mu \nu \alpha \beta}$ being the 4D anti-symmetric Levi-Civita symbol and 
$F^{\alpha \beta} = \partial ^\alpha A^\beta - \partial ^\beta A^\alpha$ is the Faraday field strength tensor. 
The term $S^\nu$ is the axial spin 4-vector (see the discussion in \cite{jackson}) which is given as
\begin{equation}
\label{4-spin}
S^\mu = \left( \gamma {\vec \beta} \cdot {\vec s} ~ , ~ {\vec s} + \frac{\gamma ^2}{1+\gamma}({\vec \beta} \cdot {\vec s})
{\vec \beta} \right) = (S^0 , {\vec S}) ~,
\end{equation}
with ${\vec s} = \frac{1}{2} {\vec \sigma}$ the standard 3-vector spin and $\gamma$, ${\vec \beta}$ the usual
relativistic gamma and beta factors. In the low velocity limit $S^\nu \rightarrow (0, {\vec s})$, one can show that 
$\Delta \alpha _{AC-B}$ from \eqref{phase-ACB} reduces to the sum of $\Delta \alpha _{AC}$ and
$\Delta \alpha _{B}$ from \eqref{phase-AC} and \eqref{phase-B}. Using $S^\nu \rightarrow (0, {\vec s})$
one can split \eqref{phase-ACB} into spatial and time integrals as
\begin{equation}
\label{phase-ACB2}
\Delta \alpha _{AC-B} =  \mu  \int \left( {\cal F} _{i j} s^j dx ^i + {\cal F} _{0 j} s^j dx ^0 \right) ~.
\end{equation}
Where in the low velocity limit the 4-vector spin has become just the ordinary 3-vector spin. 
Combining this with the dual field strength tensor time-space component -- ${\cal F} _{0 j} = B_j$ --
one finds that in the low velocity limit the integrand of the second term in \eqref{phase-ACB2} becomes $\mu {\cal F} _{0 j} s^j
\rightarrow {\vec \mu} \cdot {\vec B}$ where ${\vec \mu} = \mu {\vec s}$. Next for the dual field strength tensor the space-space
components are ${\cal F} _{i j} =  \epsilon _{ijk} E_k$. Combining this with low velocity limit for $S^\nu$ the integrand of 
the first term in \eqref{phase-ACB2} becomes $\mu {\cal F} _{i j} s^j \rightarrow  \mu \epsilon _{ijk} s^j E_k \rightarrow  
\mu {\vec s} \times {\vec E} \rightarrow - {\vec E} \times {\vec \mu}$, which is just the integrand of the Aharonov-Casher phase in
\eqref{phase-AC}.

The above development is a motivation that the expression given in \eqref{phase-ACB} is the covariant generalization
of the Aharonov-Casher phase \eqref{phase-AC} and the Bernstein phase \eqref{phase-B}. It is possible to obtain this covariant
phase more rigorously. As in the original work of Aharonov and Casher we start with the Dirac equation for a neutral 
particle non-minimally coupled to via its magnetic moment to the electromagnetic field, $F_{\mu \nu}$
\begin{equation}
\label{dirac-mm}
{\cal L} = {\bar \Psi} \left( i \gamma^ \mu \partial _\mu - \frac{\mu}{2} F_{\mu \nu} \sigma ^{\mu \nu} -m \right) \Psi ~,
\end{equation}
where $m$ is the mass of the particle, $\mu$ is the magnitude of its magnetic moment, $\gamma^\mu$
are Dirac matrices and $\sigma ^{\mu \nu} = \frac{i}{2} [ \gamma^\mu , \gamma^\nu ]$. Next we define the spin projection operators
\begin{equation}
\label{spin-pro}
\Sigma _{\pm} ({\vec S}) = \frac{1}{2} \left( 1 \pm \gamma^5 \gamma ^\mu S_\mu \right) ~.
\end{equation}
This operator projects out the $+{\vec S}$ and $-{\vec S}$ components of the spinor $\Psi$ along the direction
${\vec S}$ via the expressions 
\begin{equation}
\label{proj-psi}
\Psi _{\pm} = \Sigma _{\pm} \Psi ~~~;~~~ {\bar \Psi} _{\pm} = \Sigma _{\pm} {\bar \Psi} ~.
\end{equation} 
Thus the spinor $\Psi$ can be decomposed as $\Psi = \Psi _+ + \Psi _-$. The Lagrangian in \eqref{dirac-mm} can now be
written as
\begin{eqnarray}
\label{dirac-mm-1}
{\cal L} &=& \left( {\bar \Psi}_+ + {\bar \Psi}_- \right) 
\left( i \gamma^ \mu \partial _\mu - \frac{\mu}{2} F_{\mu \nu} \sigma ^{\mu \nu} -m \right) 
\left(\Psi _+ + \Psi _- \right) \nonumber \\
{\cal L}_\pm &=& {\bar \Psi}_\pm 
\left( i \gamma^ \mu \partial _\mu - \frac{\mu}{2} F_{\mu \nu} \sigma ^{\mu \nu} -m \right) \Psi _\pm ~.
\end{eqnarray}
In the last line we note that the Lagrangian has split into two forms -- ${\cal L}_+$ or ${\cal L}_-$ -- depending
on if one has ${\bar \Psi}_+ \left( \cdots \right) {\bar \Psi} _+$ {\it or}
${\bar \Psi}_- \left( \cdots \right) {\bar \Psi} _-$. This split occurs since for the canonical Aharonov-Casher set-up the particle
beam is polarized so that one has $\Psi _+$ {\it or} $\Psi _-$. This is the reason that in the Lagrangians
in \eqref{dirac-mm-1} does not have mixed terms like ${\bar \Psi}_+ \left( \cdots \right) {\bar \Psi} _-$ since the beam is
either $\Psi _+$ or $\Psi_-$. Thus the Lagrangian in \eqref{dirac-mm-1} has been split into two separate Lagrangians -- 
one for $\Psi _+$ ({\it i.e.} ${\bar \Psi}_+ \left( i \gamma^ \mu \partial _\mu - \frac{\mu}{2} F_{\mu \nu} 
\sigma ^{\mu \nu} -m \right) \Psi _+$) and one for $\Psi _-$ ({\it i.e.}
${\bar \Psi}_- \left( i \gamma^ \mu \partial _\mu - \frac{\mu}{2} F_{\mu \nu} \sigma ^{\mu \nu} -m \right) \Psi _-$). 

We now want to work on the terms from \eqref{dirac-mm-1} of the form 
${\bar \Psi}_\pm  \left(\frac{\mu}{2} F_{\mu \nu} \sigma ^{\mu \nu} \right) \Psi _\pm$.
First, we use the standard properties of Dirac matrices and the antisymmetry of $F_{\mu \nu}$ and
$\sigma^{\mu \nu}$ to write $F_{\mu \nu} \sigma ^{\mu \nu} =i F_{\mu \nu} \gamma^\mu \gamma^\nu$. Next we note that
\begin{equation}
\label{Sigma-ac}
\gamma^\mu \gamma ^\nu = \left\{ \Sigma _\pm , \gamma^ \mu \gamma ^\nu \right\} \mp \eta ^{\mu \nu} \gamma^5 \gamma^\alpha S_\alpha
\pm i \epsilon ^{\alpha \beta \mu \nu} \gamma_\beta S_\alpha ~. 
\end{equation}
Since we are interested in the term 
${\bar \Psi}_\pm  \left(\frac{\mu}{2} F_{\mu \nu} \sigma ^{\mu \nu} \right) \Psi _\pm \rightarrow 
{\bar \Psi}_\pm  \left(i \frac{\mu}{2} F_{\mu \nu} \gamma^\mu \gamma^\nu \right) \Psi _\pm$ we can
drop $\eta ^{\mu \nu} \gamma^5 \gamma^\alpha S_\alpha$ from \eqref{Sigma-ac} since $F_{\mu \nu}$ is antisymmetric and $\eta ^{\mu \nu}$ 
is symmetric. For the remaining terms, remembering $\Psi _\pm = \Sigma _\pm \Psi$, we apply $\Sigma _\pm$ to the left and right of 
$\gamma^\mu \gamma^\nu$ above to get
\begin{eqnarray}
\label{gg-1}
\Sigma _\pm \gamma^\mu \gamma ^\nu \Sigma _\pm &=& \Sigma _\pm \left\{ \Sigma _\pm , \gamma^ \mu \gamma ^\nu \right\} \Sigma _\pm
\pm i \Sigma _\pm \epsilon ^{\alpha \beta \mu \nu} \gamma_\beta S_\alpha \Sigma _\pm \nonumber \\
&=& 2 \Sigma _\pm \gamma^\mu \gamma ^\nu \Sigma _\pm
\pm i \Sigma _\pm \epsilon ^{\alpha \beta \mu \nu} \gamma_\beta S_\alpha \Sigma _\pm \\
&=& \mp i \Sigma _\pm \epsilon ^{\alpha \beta \mu \nu} \gamma_\beta S_\alpha \Sigma _\pm 
\nonumber ~,
\end{eqnarray}
where in the second line we have used the fact that $\Sigma_ \pm$ is a projection operator {\it i.e.}
$\Sigma_ \pm \Sigma_ \pm = \Sigma_ \pm$. Putting all this together we can write the Lagrangians in \eqref{dirac-mm-1} as 
\begin{equation}
\label{dirac-mm-3}
{\cal L}_\pm = {\bar \Psi}_\pm \left( \gamma^\mu [i \partial _\mu - \mu {\cal F} _{\mu \nu} S^\nu ] -m \right) \Psi _\pm ~,
\end{equation}
The Lagrangian of the usual minimally coupled Dirac particle with charge $e$  has the form 
${\cal L}_{Dirac} = {\bar \Psi} \left( \gamma^\mu [i \partial _\mu - e A_\mu ] -m \right) \Psi$ with $A_\mu$ being the 
4-vector potential. Comparing this Dirac Lagrangian with those in \eqref{dirac-mm-3} we find the correspondence 
$e A _\mu \leftrightarrow \mu {\cal F} _{\mu \nu} S^\nu$ (this correspondence was also noted in
\cite{dulat}). Then using the fact that the minimal coupling $i \partial _\mu - e A_\mu$  gives the covariant Aharonov-Bohm 
phase as $e \oint A_\mu dx^\mu$ we conclude that the coupling in \eqref{dirac-mm-3}, $i \partial _\mu - \mu {\cal F} _{\mu \nu} S^\nu $  
gives the covariant Aharonov-Casher phase 
\begin{equation}
\label{phase-ACB-3}
\Delta \alpha _{AC-B} =  \mu \int {\cal F} _{\mu \nu} S^\nu dx ^\mu ~.
\end{equation}
This agrees with the previously given form in \eqref{phase-ACB}. However here we have derived the covariant Aharonov-Casher
phase starting from the Lagrangian in \eqref{dirac-mm}, whereas previously we only gave a motivation for the covariant 
form of the AC phase via heuristic arguments starting from the 3-vector form of the Aharonov-Casher phase
({\it i.e.} - $\mu \int _L ( {\vec E} \times {\vec s} )  \cdot d {\vec r}$) and the
3-vector form of the Bernstein/scalar Aharonov-Bohm phase ({\it i.e.} $\int _L ( {\vec \mu} \cdot {\vec B}  ) dt$).

We now use the expression in \eqref{phase-ACB2} to make two comments about the combined Aharonov-Casher 
and Bernstein effect. 

First, from the \eqref{phase-ACB2} we can write the integrand as a total differential 
$$
d(\Delta \alpha _{AC-B}) = \partial _i (\Delta \alpha _{AC-B} ) dx^i + \partial_0 (\Delta \alpha _{AC-B}) dx^0
=  \mu \left( {\cal F} _{i j} s^j dx ^i + {\cal F} _{0 j} s^j dx ^0 \right) ~,
$$
which gives the identities $\partial _i (\Delta \alpha _{AC-B} ) = \mu {\cal F} _{i j} s^j$ and
$\partial_0 (\Delta \alpha _{AC-B}) =  \mu {\cal F} _{0 j} s^j$. Now by the equality of mixed partial 
derivatives $\partial _0 \partial _i  (\Delta \alpha _{AC-B}) = \partial _i \partial _0  (\Delta \alpha _{AC-B})$
we have
\begin{equation}
\label{force}
\partial _0 (\mu {\cal F} _{i j} s^j ) = \partial _i (\mu {\cal F} _{0 j} s^j )~.
\end{equation}
Using ${\cal F} _{i j} s^j =  \epsilon _{ijk} E_k s^j = - {\vec E} \times {\vec s}$ and
$\mu {\cal F} _{0 j} s^j = {\vec B} \cdot {\vec s}$. Plugging this into \eqref{force} yields
\begin{equation}
\label{force2}
\frac{1}{c} \frac{d}{dt} ( {\vec E} \times {\vec s} ) = - \nabla ({\vec B} \cdot {\vec s} )~.
\end{equation} 
In \eqref{force2} we were able to replace the partial time derivative by the full time derivative since
$$
\frac{d}{dt} ( {\vec E} \times {\vec s} ) = \frac{d{\bf x}}{dt} \cdot \left[\nabla ({\vec E} \times {\vec s} )\right]
+ \frac{\partial}{\partial t} ( {\vec E} \times {\vec s} ) \approx \frac{\partial}{\partial t} ( {\vec E} \times {\vec s} ) ~,
$$ 
since in the low velocity limit $\frac{d{\bf x}}{dt} \approx 0$. The result in \eqref{force2} implies that the force
on the neutral magnetic moment, ${\vec \mu} = \mu {\vec s},$ moving in the combined electric and magnetic field is zero
in the low velocity limit. \footnote{From reference \cite{yarman} the 3-vector expression for the force on a magnetic 
dipole ${\vec \mu}$ moving in a combined electric and magnetic field is given by ${\vec F} = \nabla ({\vec \mu} \cdot {\vec B}) 
- \frac{1}{c}\frac{d ({\vec \mu} \times {\vec E})}{dt}$ so that \eqref{force2} implies ${\vec F} = 0$ in this low velocity limit.} 
The vanishing of the classical force on the particle is one of the conditions of the AB and AC effects.

Second, the canonical Aharonov-Casher effect as given by the phase in \eqref{phase-AC} is non-dispersive {\it i.e.}
the phase shift is independent of the velocity of the particle. This same non-dispersive feature holds 
for the Aharonov-Bohm effect as well. This is a crucial aspect of the Aharonov-Bohm effect since if a charged particle
interacts directly with a magnetic field, ${\vec B}$, one will also in general get phase shifts due to the 
${\vec v} \times {\vec B}$ forces on the particle. However the phase shifts associated with the 
${\vec v} \times {\vec B}$ forces are explicitly velocity dependent and thus dispersive. 
Now for the covariant expression of the AC phase given in \eqref{phase-ACB} or \eqref{phase-ACB2} we see that,
if we do not make the low velocity limit, that the AC phase shift will depend on the velocity ${\vec \beta}$, 
and will in general be dispersive. Thus the non-dispersive nature of the Aharonov-Casher effect is a result of
taking the low velocity limit. For the non-covariant forms of the Aharonov-Bohm phase of \eqref{phase-AB} or 
the Aharonov-Casher phase of \eqref{phase-AC}, 
the non-dispersive character of these phases is one of their fundamental features. From the covariant expression 
of the phase in equation \eqref{phase-ACB} one finds that there will in general be a velocity dependence due to the 
form of the $4$-spin in \eqref{4-spin}. Thus in general the AC effect is dispersive. In contrast the covariant AB phase
given by $e \oint A_\mu dx^\mu$ is still independent of the particle velocity and is non-dispersive even if one relaxes the 
low velocity limit.

In the next section we will examine the case where a neutral particle with a magnetic moment moves in the background field 
of a plane electromagnetic wave. The plane wave has both electric and magnetic fields, and both fields lead to non-zero phases
which have an interesting interplay with one another. 

\section{Aharonov-Casher-Bernstein effect in a plane electromagnetic wave background}
\label{t-depend}
We now want to use the results of the previous section to study a specific example of 
the time-dependent Aharonov-Casher-Bernstein effect. In the original Aharonov-Casher paper and
in the original Bernstein paper the electric and magnetic fields were static. For example,
the original Aharonov-Casher setup was for the static electric field of an infinite line charge.
If this line charge were allowed to varying in time then one would generate magnetic fields
which would lead to additional phase shifts and therefore the simple 3-vector expression for the 
Aharonov-Casher phase, equation \eqref{phase-AC}, would not be correct; one would need a covariant
expression, such as our proposal in equation \eqref{phase-ACB2}, in order to handle the additional
phase shift coming from the time dependence of the fields. 

The particular time varying electric and magnetic background we consider is a 
plane electromagnetic wave traveling in the $+z$ direction and polarized along the $x$ direction. The 
fields for this are
\begin{equation}
\label{em}
{\vec E} = f(\omega t - k z) {\hat {\bf x}} ~~~~; ~~~~ {\vec B} = f(\omega t - k z) {\hat {\bf y}}
\end{equation}
We have written the electric and magnetic fields in terms of arbitrary wave forms, but for concreteness 
one can take them to be sinusoidal {\it e.g}  $f(\omega t - k z) = E_0 \sin (\omega t - k z)$.
We now use the covariant phase shift expression given in \eqref{phase-ACB2} to investigate the phase shift on
a neutron moving in the background of the linearly polarized plane wave of \eqref{em}. There are three
choices for the direction of ${\vec s}$ (and therefore the direction of ${\vec \mu}$): ${\hat {\bf x}}$, 
${\hat {\bf y}}$, or ${\hat {\bf z}}$. Choosing ${\vec s} \sim {\hat {\bf x}}$ gives  ${\vec E} \times {\vec s} =0$ 
and ${\vec B} \cdot {\vec s} = 0$ so we would get no additional phase shift for this
case. If ${\vec s} \sim {\hat {\bf z}}$ then ${\vec B} \cdot {\vec s} = 0$ while 
${\vec E} \times {\vec s} \sim {\hat {\bf y}}$. In this case, if the particle follows a path along the $y$ direction 
with $d{\vec r} \sim {\hat {\bf y}}$,  one gets a time varying AC phase shift from the electric field, but no 
Bernstein phase shift. The details of this time varying AC phase shift will depend on the details of the velocity of the particle
and the frequency of the wave , $\omega$. The case ${\vec s} \sim {\hat {\bf y}}$ is the most interesting and most general 
since now both the Aharonov-Casher and Bernstein phase shifts are non-zero since
${\vec E} \times {\vec s} \ne 0$ and ${\vec B} \cdot {\vec s} \ne 0$. In this case one finds an interesting 
interplay between the two effects.

For the electromagnetic fields given by \eqref{em} and the spin in the direction
${\vec s} \sim {\hat {\bf y}}$, the combined Aharonov-Casher-Bernstein phase shift according to \eqref{phase-AC} 
\eqref{phase-B} and \eqref{phase-ACB2} for the particle traveling with velocity $v$ in the $\pm z$-direction is
(we restore a factor of $\frac{1}{c}$)
\begin{eqnarray}
\label{em-phase}
\Delta \alpha _{AC-B} &=&  \frac{\mu}{c} \int {\cal F} _{\mu \nu} S^\nu dx ^\mu = 
 \frac{\mu}{c} \int _L ( {\vec s} \cdot {\vec B}  ) c~ dt - \frac{\mu}{c} \int _L ( {\vec E} \times {\vec s} )  \cdot d {\vec r} \nonumber \\
&=& \frac{\mu }{2v} \int_L  [f(\omega t - k z)] dz - \frac{\mu}{2c} \int _L [f(\omega t - k z)] {\hat {\bf z}} \cdot (\pm dz {\hat {\bf z}}) \\
&=&  \frac{\mu}{2c} \int_0 ^L [f(\omega t - k z)] \left(\frac{c}{v} \mp 1\right) dz ~. \nonumber
\end{eqnarray}
In the second line we have used $dt = dz /|{\bf v}|$ and the $\pm$ in the electric field contribution indicates if the particle 
is going along $\pm {\hat {\bf z}}$. The factor of $\frac{1}{2}$ comes from the spin half of the particle.
Note that when the particle moves with the wave ({\it i.e.} for
particles moving in the $+{\hat {\bf z}}$ direction) the magnetic and electric effects tend to cancel. This can be seen
from the factor of $\frac{c}{v}-1$ in \eqref{em-phase} which goes to zero as the particle velocity approaches $c$ {\it i.e} $v \sim c$. 
(However in this ultra-relativistic limit we would need to take into account the full 4-vector spin, $S^\mu$, rather than simply
the 3-vector spin , ${\vec s}$, as we did in \eqref{em-phase}).
When the particle moves against the direction of the wave ({\it i.e.} for particles moving in the - ${\hat {\bf z}}$ direction) 
the magnetic and electric effects tend to add. This can be seen from the factor of 
$\frac{c}{v}+1$ in \eqref{em-phase} which goes to $2$ as the particle velocity approaches $c$ {\it i.e} $v \sim c$.
This cancellation or adding, depending on the direction of travel of the particle in comparison
to that of the wave, was also found in the case of the time-dependent Aharonov-Bohm for plane waves \cite{bright}.  

We now use \eqref{em-phase} to calculate the total Aharonov-Casher-Bernstein phase shift for two particles starting at 
$z=0$ and $t=0$ going out to $z=\pm \Delta z$ in time $t=\Delta t$, then turning around and returning to $z=0$ at time
$t=2 \Delta t$. The magnitude of the velocity of the particles is $|{\bf v}| = \frac{\Delta z}{\Delta t}$. The {\it closed} space-time 
path of the particles is shown in figure 1. One particle moves along paths $1 + 2$ and the other particle moves along the
paths $3+4$. To get a closed space-time path we reverse the direction of the particle going along paths $3+4$ by multiplying the
line integrals for these paths by a minus sign and then adding them to the line integrals from paths $1+2$. This is indicated in
the figure by the reversal of the arrows along paths $3+4$.

\begin{figure}
  \centering
	\includegraphics[trim = 0mm 0mm 0mm 0mm, clip, width=6.0cm]{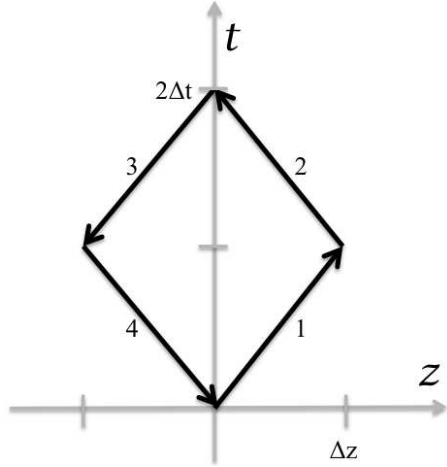}
\caption{{\it The closed space-time loop in the $t-z$ plane}}
\label{fig1}
\end{figure}

{\bf Path 1 :} For path 1 we have $z = v t$ and $\Delta z = v \Delta t$ so that the argument of the function $f$ becomes
$\omega t - k z = \frac{\omega}{v} z - k z = k \left(\frac{c}{v} -1 \right) z$, where we have used $\frac{\omega}{k}=c$. Now defining
$k_- = k \left(\frac{c}{v} -1 \right)$ the line integral for path 1 from \eqref{em-phase} is
\begin{equation}
\label{path1}
\frac{\mu}{c} \int _1 {\cal F} _{\mu \nu} S^\nu dx ^\mu = \frac{\mu}{2c} \left(\frac{c}{v} - 1\right) \int_0 ^{\Delta z} f(k_- z)  dz ~.
\end{equation}
Now defining $F(\zeta) = \int f(\zeta) d \zeta$ ({\it e.g.} if $f(\zeta) = E_0 \sin (\zeta)$ then 
$F(\zeta) = - E_0 \cos (\zeta) + K$) \eqref{path1} integrates to
\begin{equation}
\label{path1a}
\frac{\mu}{c} \int _1 {\cal F} _{\mu \nu} S^\nu dx ^\mu = \frac{\mu}{2 k_- c} \left(\frac{c}{v} - 1\right) [F(k_- \Delta z) - F(0)]
= \frac{\mu}{2 \omega} [F(k_- \Delta z) - F(0)] ~,
\end{equation}
where we have used $k_- = k \left(\frac{c}{v} -1 \right)$ and $kc= \omega$.

{\bf Path 2 :} For path 2 we have $z = - v t + 2 \Delta z$ and $\Delta t < t < 2 \Delta t$ so that $z$ goes from $z=\Delta z$
to $z=0$. The argument of the function $f$ becomes $\omega t - k z = 2 \Delta k \frac{c}{v} - k \left(\frac{c}{v} + 1 \right) z$. Now defining
$k_+ = k \left(\frac{c}{v} + 1 \right)$ the line integral for path 3 from \eqref{em-phase} is
\begin{equation}
\label{path2}
\frac{\mu}{c} \int _2 {\cal F} _{\mu \nu} S^\nu dx ^\mu = \frac{\mu}{2c} \left(\frac{c}{v} + 1\right) \int ^0 _{\Delta z} 
f \left(2 \Delta k \frac{c}{v} - k_+ z \right)  dz ~.
\end{equation}
Integration of \eqref{path2} leads to
\begin{equation}
\label{path2a}
\frac{\mu}{c} \int _2 {\cal F} _{\mu \nu} S^\nu dx ^\mu = - \frac{\mu}{2 \omega} \left[ F \left( 2 k \Delta z \frac{c}{v} \right) - F(k_- \Delta z) \right] ~,
\end{equation}
Note that while the integral in \eqref{path2} begins with $k_+$ the final result has $k_-$.

{\bf Path 3 :} For path 3 we have $z = v t - 2 \Delta z$ and $\Delta t < t < 2 \Delta t$ so that $z$ goes from $z=-\Delta z$
to $z=0$. The argument of the function $f$ becomes $\omega t - k z = 2 \Delta k \frac{c}{v} + k \left(\frac{c}{v} - 1 \right) z$. Recalling
the definition $k_- = k \left(\frac{c}{v} - 1 \right)$ the line integral for path 3 from \eqref{em-phase} is
\begin{equation}
\label{path3}
\frac{\mu}{c} \int _3 {\cal F} _{\mu \nu} S^\nu dx ^\mu = \frac{\mu}{2c} \left(\frac{c}{v} - 1\right) \int ^0 _{-\Delta z} 
f \left(2 \Delta k \frac{c}{v} + k_- z \right)  dz ~.
\end{equation}
Integration of \eqref{path3} leads to
\begin{equation}
\label{path3a}
\frac{\mu}{c} \int _3 {\cal F} _{\mu \nu} S^\nu dx ^\mu = \frac{\mu}{2 \omega} \left[ F \left( 2 k \Delta z \frac{c}{v} \right) - F(k_+ \Delta z) \right] ~,
\end{equation}
The integral in \eqref{path3} begins with $k_-$ the final result has $k_+$. A similar thing happened in the integration of path 2.

{\bf Path 4 :} Finally, for path 4 we have $z = - v t$ and $\Delta z = - v \Delta t$ so that the argument of the function $f$ becomes
$\omega t - k z = \frac{\omega}{c} z - k z = k \left(\frac{c}{v} +1 \right) z$, where we have used $\frac{\omega}{k}=c$. Recalling
$k_+ = k \left(\frac{c}{v} +1 \right)$ the line integral for path 4 from \eqref{em-phase} is
\begin{equation}
\label{path4}
\frac{\mu}{c} \int _4 {\cal F} _{\mu \nu} S^\nu dx ^\mu = \frac{\mu}{2c} \left(\frac{c}{v} + 1\right) \int_0 ^{-\Delta z} f(-k_+ z)  dz ~.
\end{equation}
Integration of \eqref{path4} leads to
\begin{equation}
\label{path4a}
\frac{\mu}{c} \int _4 {\cal F} _{\mu \nu} S^\nu dx ^\mu = - \frac{\mu}{2 k_+c} \left(\frac{c}{v} + 1\right) [F(k_+ \Delta z) - F(0)]
= \frac{\mu}{2 \omega} [F(k_+ \Delta z) - F(0)] ~,
\end{equation}
where we have used $k_+ = k \left(\frac{c}{v} -1 \right)$ and $kc = \omega$.

We now combine the results from \eqref{path1a} \eqref{path2a} \eqref{path3a} \eqref{path4a} (remembering to put a minus 
sign in front of the results from path 3 and path 4 so that we get a closed space-time path) to get
\begin{eqnarray}
\label{opath}
\frac{\mu}{c} \oint {\cal F} _{\mu \nu} S^\nu dx^\mu &=& \frac{\mu}{c} \left( \int _1 {\cal F} _{\mu \nu} S^\nu dx^\mu + 
\int _2 {\cal F} _{\mu \nu} S^\nu dx^\mu 
+ \int _3 {\cal F} _{\mu \nu} S^\nu dx^\mu + \int _4 {\cal F} _{\mu \nu} S^\nu dx^\mu \right) \nonumber \\
&=& \frac{\mu}{\omega} \left[ -F(0) - F \left( 2 k \Delta z \frac{c}{v} \right) + F(k_+ \Delta z) + F (k_- \Delta z) \right] ~,
\end{eqnarray}
where $k_\pm = \left(1 \pm \frac{c}{v} \right)$. Expanding the integration functions to second order 
$F(\eta) = F(0) + \eta F'(0) + \frac{1}{2} \eta ^2 F''(0) +...$ gives for the loop integral in \eqref{opath}
\begin{equation}
\label{opath-2}
\frac{\mu}{c} \oint {\cal F} _{\mu \nu} S^\nu dx^\mu \approx \frac{\mu}{\omega} \left[ F''(0) k^2 \Delta z ^2 \left( 1- \frac{c^2}{v^2} \right)\right] ~.
\end{equation}
The zero and first order terms cancel so that one finds that $\frac{\mu}{c} \oint {\cal F} _{\mu \nu} S^\nu dx^\mu$ 
is non-zero only starting at second order. A similar result was found in the case of the Aharonov-Bohm phase for a 
plane wave background \cite{bright}. This result can be attributed to the interplay and partial cancellation between 
the electric and magnetic field contribution to the phases.  

\section{Summary and Conclusions}
Here we have given a covariant expression for the Aharonov-Casher effect in equations \eqref{phase-ACB} \eqref{phase-ACB-3} as a 
generalization of the usual non-covariant 3-vector expression from \eqref{phase-AC}. This covariant expression also includes 
the Bernstein or scalar Aharonov-Bohm phase shift. In 3-vector, non-covariant 
form the Bernstein phase is given in equation \eqref{phase-B}. One encounters
the Bernstein/scalar Aharonov-Bohm effect when a neutral particle with a magnetic dipole moment moves through a magnetic field. 
One conclusion of the of the covariant expression \eqref{phase-ACB2} is that in general the Aharonov-Casher phase is
dispersive {\it i.e.} depends on the velocity of the particle. This velocity dependence comes in through the form
of the 4-vector spin, $S^\mu$, given in \eqref{4-spin} which depends on ${\vec \beta}$. In contrast one of the hallmarks of the Aharonov-Bohm
phase, $\frac{e}{\hbar c} \oint A_\mu dx^\mu$, is its non-dispersive character. As well the standard non-covariant,
3-vector Aharonov-Casher phase of \eqref{phase-AC} is velocity independent/non-dispersive. 

The covariant expression of \eqref{phase-ACB} \eqref{phase-ACB-3} can be used to analyze situations where one has both electric and 
magnetic fields such as occurs generally in time dependent situations. Here we used the covariant expression to investigate
the phase shift that occurs when a neutron moves in the background field of a plane, linearly polarized electromagnetic wave.
The final result of the phase shift for this kind of background (and for the diamond path shown in figure 1) is that 
the time-dependent Aharonov-Casher phase vanishes to first order -- from \eqref{opath-2} we see that the first non-zero
term comes from the second order term in the expansion of $F(\zeta )$. This approximate vanishing can be attributed to
the interplay and partial cancellation between the electric and magnetic contributions to the phase. A similar partial
cancellation between the electric and magnetic contributions to the phase has been seen in the case of the time-dependent 
Aharonov-Bohm effect \cite{singleton, singleton2, bright}. \\

{\par\noindent {\bf Acknowledgments:}} DS is supported by a 2015-2016 Fulbright Scholars Grant to Brazil. 
DS wishes to thank the ICTP-SAIFR in S{\~a}o Paulo for it hospitality. DS also acknowledges support by a grant 
(number 1626/GF3) in Fundamental Research in Natural Sciences by the Science Committee of the Ministry of Education 
and Science of Kazakhstan. \\

\end{document}